\documentclass[12pt]{article}

\usepackage{color}
\usepackage{graphicx}
\usepackage{amsfonts}
\usepackage{amsthm}
\usepackage{multirow}
\usepackage[top=3cm, bottom=3.25cm, left=2.75cm, right=2.75cm]{geometry}

\newcommand{\Section}[1]{\section{#1} \setcounter{equation}{0}}
\newcommand{\beq}{\begin{equation}}
\newcommand{\eeq}[1]{\label{#1}\end{equation}}
\newcommand{\ber}{\begin{eqnarray}}
\newcommand{\eer}[1]{\label{#1}\end{eqnarray}}

\newcommand{\cdt}{\!\cdot\!}

\newcommand{\ft}[2]{{\textstyle\frac{#1}{#2}}}
\newcommand{\nll}{N\!=\!(1,1)}
\newcommand{\nZZ}{N\!=\!(2,2)}

\newcommand{\myF}{\mathrm{F}}
\newcommand{\ff}{\myF}
\newcommand{\ffb}{{\bar\myF}}
\newcommand{\ffh}{\hat{\myF}}
\newcommand{\ffhb}{{\hat{\bar\myF}}}
\newcommand{\fft}{{\tilde\myF}}
\newcommand{\fftb}{{\bar{\tilde\myF}}}
\newcommand{\ffth}{{\hat{\tilde\myF}}}
\newcommand{\fftbh}{\hat{\bar{\tilde\myF}}}
\newcommand{\ffs}{\mathbb{F}}
\newcommand{\ffbs}{{\bar\mathbb{F}}}
\newcommand{\ffts}{{\tilde\mathbb{F}}}
\newcommand{\fftbs}{{\bar{\tilde\mathbb{F}}}}

\newcommand{\vv}[1]{\mathbb{V}^{#1}}
\newcommand{\vvb}[1]{\bar{\mathbb{V}}^{#1}}
\newcommand{\vvt}[1]{\tilde{\mathbb{V}}^{#1}}
\newcommand{\vvtb}[1]{\bar{\tilde{\mathbb{V}}}^{#1}}

\newcommand{\dt}[1]{\hat{d}^{#1}}
\newcommand{\dq}[1]{q^{#1}}
\newcommand{\dqh}[1]{\hat{q}^{#1}}
\newcommand{\jj}[3]{{J}_{#1}{}^{#2}{}_{#3}}
\newcommand{\bbD}[1]{\mathbb{D}_{#1}}
\newcommand{\bbDB}[1]{\bar{\mathbb{D}}_{#1}}
\newcommand{\bbG}[1]{\mathbb{G}_{#1}}
\newcommand{\bbGB}[1]{\bar{\mathbb{G}}_{#1}}
\newcommand{\bbX}[1]{\mathbb{X}_{#1}}
\newcommand{\bbXB}[1]{\bar{\mathbb{X}}_{#1}}

\newcommand{\hee}[1]{\Xi^1_{#1}}
\newcommand{\hcc}[1]{\Xi^2_{#1}}

\def\+{{+\!\!\!+}}

\newcommand{\del}{\delta}
\newcommand{\aleq}{&\!\!\!=\!\!\!&}
\newcommand{\nn}{\nonumber}
\newcommand{\kah}{K\"ahler~}

\newcommand{\lam}{\Lambda}

\newcommand{\lamb}{\bar\Lambda}
\newcommand{\lamt}{\tilde\Lambda}
\newcommand{\lamtb}{\bar{\tilde{\Lambda}}}

\newcommand{\eg}{\textit{e.g.},~}

\def\+{{+\!\!\!+}}

\begin{document}
\renewcommand{\theequation}{\thesection.\arabic{equation}}
\setcounter{page}{0}
\thispagestyle{empty}

\begin{flushright} \small
UUITP-11/08 \\  YITP-SB-08-29\\
\end{flushright}
\smallskip
\begin{center} \LARGE
{\bf  Nonabelian Generalized Gauge Multiplets}
 \\[12mm] \normalsize
{\bf Ulf~Lindstr\"om$^{a}$, Martin Ro\v cek$^{b}$, Itai Ryb$^{b}$,\\
Rikard von Unge$^{b,c,d}$, and Maxim Zabzine$^{a}$} \\[8mm]
{\small\it
$^a$Department of Theoretical Physics
Uppsala University, \\ Box 803, SE-751 08 Uppsala, Sweden \\
~\\
$^b$C.N.Yang Institute for Theoretical Physics, Stony Brook University, \\
Stony Brook, NY 11794-3840,USA\\
~\\
$^c$Simons Center for Geometry and Physics, Stony Brook University, \\
Stony Brook, NY 11794-3840,USA\\
~\\
$^{d}$Institute for Theoretical Physics, Masaryk University, \\
61137 Brno, Czech Republic \\~\\}
\end{center}
\vspace{10mm}
\centerline{\bf\large Abstract}
\bigskip
\noindent
We give the nonabelian extension of the newly discovered $\nZZ$ two-dimensional vector multiplets. These can be used to gauge symmetries of sigma models on generalized \kah geometries. Starting from the transformation rule for the nonabelian case we find covariant derivatives and gauge covariant field-strengths and write their actions in $\nZZ$ and $\nll$ superspace.
\eject
\normalsize

\eject

\Section{Introduction}
Studying nonlinear sigma models whose target spaces are generalized 
\kah geometries in $\nZZ$ superspace, we found new vector multiplets 
\cite{Lindstrom:2007vc,Gates:2007ve,Lindstrom:2007sq,ryb} that can 
be used to gauge isometries that act on different types of superfields 
\cite{Grisaru:1997ep}. In this note, we extend these results to the nonabelian case; 
in particular we find the algebra of the gauge-covariant superspace derivatives. 
Our results make possible new nonabelian gaugings and quotient constructions,
and give insight into the geometric meaning of the new multiplets. 

The plan of the paper is as follows: In the next section,
we review the abelian multiplets \cite{Lindstrom:2007vc,Gates:2007ve}. In section 3, we
discuss the nonabelian extensions of the large vector multiplet, 
which couples chiral and twisted chiral gauge symmetries \cite{ggg}; we give the fundamental
superfield gauge potentials, construct covariant derivatives as well
as field strengths in $\nZZ$ superspace, reduce to $\nll$ superspace and 
discuss actions ({\it cf.}\ \cite{ryb} for the abelian case). In
section 4, we repeat this discussion for the semichiral vector multiplet \cite{bbb}. 
We end with a few remarks.
We follow the notation of \cite{Lindstrom:2007sq}. 

\Section{Abelian vector multiplets}
Until recently, two $\nZZ$ vector multiplets were known. Both are described by a single unconstrained scalar superfield $V$ and differ by their gauge transformations: The chiral gauge multiplet transforms with a chiral gauge parameter ($\bbDB{\pm} \lam=0$)
\beq
\del V^\phi = i(\lamb-\lam)~,
\eeq{vf}
whereas the {\em twisted} chiral gauge multiplet transforms with a twisted chiral gauge parameter ($ \bbDB{+} \lamt = \bbD{-}\lamt=0$)
\beq
\del V^\chi = i(\lamtb-\lamt)~.
\eeq{vc}

These multiplets have gauge invariant twisted chiral and chiral field strengths, respectively:
\beq
\tilde W = \bbDB{+}\bbD{-} V^\phi~,~~~ W=\bbDB{+}\bbDB{-} V^\chi~.
\eeq{wtw}

In \cite{Lindstrom:2007vc}, we introduced two new multiplets: the large
vector multiplet (rewritten here in the conventions of \cite{Lindstrom:2007sq})
\beq
\delta V^\phi = i (\lamb-\lam) ~~,~~~
\delta V^\chi = i(\lamtb-\lamt) ~~,~~~
\delta V^\prime = (-\lam-\lamb+\lamt+\lamtb)~,
\eeq{lvm}
which transforms with chiral {\em and} twisted chiral parameters, and
the semichiral vector multiplet (see also \cite{Gates:2007ve})
\beq
\del \vv L = i (\lamb^L-\lam^L) ~~,~~~
\del \vv R = i(\lamb^R-\lam^R) ~~,~~~
\del \vv\prime = (-\lam^L-\lamb^L+\lam^R+\lamb^R)~,
\eeq{svm}
whose gauge parameters are semichiral: $\bbDB{+}\lam^L=\bbDB{-}\lam^R=0$. In both cases it is useful to introduce  complex linear combinations with simple transformations.

For the large vector multiplet, one finds the combinations
\begin{eqnarray}
V_L \aleq \ft{1}{2} (-V^\prime + i (V^\phi-V^\chi)) ~\Rightarrow~
\delta V_L = \lam-\lamt~, \nn \\ 
\label{vright} V_R \aleq \ft{1}{2} (-V^\prime + i (V^\phi+V^\chi))~ \Rightarrow~
\delta V_R = \lam-\lamtb~.
\end{eqnarray}
Note that $V_{L,R}$ are constrained, as they have the same real part
\begin{equation}
 V_L + \bar{V}_L =  V_R + \bar{V}_R~,
\end{equation}
or equivalently,
\beq
V_L=V_R-iV^\chi
\eeq{vlvrchi}
This constraint is preserved by the gauge transformations. 
The field-strengths of the large vector multiplet are semichiral spinors:
\beq
 \bbG{+} = \bbDB{+} V_L~,~~\bbG{-} = \bbDB{-} V_R
  ~,~~\bbGB{+} = \bbD{+} \bar{V}_L~,~~
  \bbGB{-} = \bbD{-} \bar{V}_R ~.
\eeq{lvmfield}

For the semichiral vector multiplet, we find similar combinations:
\beq 
\vv{}=\ft{1}{2} (-\vv\prime + i (\vv L-\vv R)) ~\Rightarrow~
\delta \vv{} = \lam_L-\lam_R~,
\eeq{semiv}
\beq
\vvt{}=\ft{1}{2} (-\vv\prime + i (\vv L+\vv R)) ~\Rightarrow~
\delta \vvt{} = \lam_L-\lamb_R~.
\eeq{semivt}
As for the large vector multiplet, these combinations are constrained to have the same real part. The field-strengths of the semichiral vector multiplet are chiral and twisted chiral scalars:
\beq
\ffs = \bbDB{+}\bbDB{-}\vv{}~,~~
\ffbs =-\bbD{+}\bbD{-}\vvb{}~,~~
\ffts = \bbDB{+}\bbD{-} \vvt{}~,~~
\fftbs = - \bbD{+}\bbDB{-}\vvtb{}~.
\eeq{semifield}

\Section{Nonabelian Large Vector Multiplet}
\subsection{Covariant derivatives and field-strengths}
The nonabelian generalizations of (\ref{vf}) and (\ref{vc}) are well known; for finite gauge transformations they are
\beq
g(\lam)e^{V^\phi}=e^{i\lamb}e^{V^\phi}e^{-i\lam}~,~~
g(\lamt)e^{V^\chi}=e^{i\lamtb}e^{V^\chi}e^{-i\lamt}~.
\eeq{nonavfc}
These clearly carry over for (\ref{lvm}), {\em except} for the transformation of $V^\prime$, which cannot easily be generalized in a way compatible with the group property. Instead, we use the complex potential $V_R$, and postulate
\beq
g(\lam,\lamt) e^{-iV_R}=e^{i\lamtb}e^{-iV_R}e^{-i\lam}~
\Rightarrow~ g(\lam,\lamt) e^{i\bar V_R}=e^{i\lamb}e^{i\bar V_R}e^{-i\lamt}~.
\eeq{nonalvmvr}
This choice is arbitrary, as we could have formulated (\ref{nonalvmvr}) using the complex potential $V_L$; 
we define it by $e^{-iV_L}=e^{-V^\chi}e^{-iV_R}$ ({\it cf.} eq.~(\ref{vlvrchi})).

To avoid introducing extra degrees of freedom, we impose a gauge covariant reality condition:
\beq
e^{i\bar V_R}=e^{V^\phi}e^{iV_R}e^{V^\chi}~,
\eeq{nonalvmreal}
which is compatible with (\ref{vright}), as it reduces to 
$i(\bar V_R-V_R)= V^\phi+V^\chi$ in the Abelian limit.

Gauging proceeds in usual way, {\it e.g.,} 
\beq
K(\bar\Phi\cdt\Phi,\bar\chi\cdt\chi,\bar\Phi\cdt\chi,\bar\chi\cdt\Phi)\to
K(\bar\Phi\, e^{V^\phi\!}\cdt\Phi,\bar\chi\, e^{V^\chi\!}\cdt\chi,\bar\Phi\, 
e^{i\bar V_{R}}\cdt\chi,\bar\chi\, e^{-iV_{R}}\cdt\Phi)
\eeq{Kgauge}
for $\Phi,\chi$ in the same (arbitrary) representation of the nonabelian gauge group.

Covariant derivatives can be constructed in different representations appropriate to the matter fields they act on. Here we start with the chiral representation, which acts naturally on chiral superfields. In this representation, the covariant derivatives $\nabla$ transform as
\beq
g(\lam) \nabla = e^{i\lam}\nabla e^{-i\lam}~.
\eeq{chiralnabla}
Other representations can be found by conjugating with the appropriate
combinations of $e^{V^\phi},e^{V^\chi}$ and $e^{iV_R}$.
Because the gauge parameter $\lam$ is chiral, $\bbDB{\pm} \lam=0$,
$\bbDB{\pm}$ are already covariant
\beq
\bar{\nabla}_\pm = \bbDB{\pm}~,
\eeq{dbar_cov}
that is, they transform as (\ref{chiralnabla}). Likewise, the usual expressions
\beq
\nabla_\pm=e^{-V^\phi}\bbD{\pm} e^{V^\phi}
\eeq{nablavf}
are covariant. However, because $\bbDB{+}\lamt=\bbD{-}\lamt=0$, there are more covariantly
transforming derivatives; a quick calculation shows that
\ber
\hat{\nabla}_+ \aleq e^{iV_R}\bbD{+}e^{-iV_R} \nn \\ 
\hat{\bar{\nabla}}_- \aleq e^{iV_R}\bbDB{-}e^{-iV_R} \nn \\
\hat{\bar{\nabla}}_+ \aleq e^{iV_R}e^{V^\chi}\bbDB{+}e^{-V^\chi} e^{-iV_R} \nn \\
\hat{\nabla}_- \aleq e^{iV_R}e^{V^\chi}\bbD{-}e^{-V^\chi} e^{-iV_R}
\eer{morenabla}
are also good covariant derivatives. The derivatives (\ref{dbar_cov}) and (\ref{nablavf})
are simple in chiral or antichiral representation. A twisted chiral representation is obtained after a similarity transformation with $e^{-iV_L}=e^{-V^\chi}e^{-iV_R}$; 
then the supercovariant derivatives $\hat \nabla$ become simple:
\ber
e^{-iV_L}\hat{\nabla}_+e^{iV_L} \aleq e^{-V^\chi}\bbD{+}e^{V^\chi} \nn \\ 
e^{-iV_L}\hat{\bar{\nabla}}_- e^{iV_L}\aleq e^{-V^\chi}\bbDB{-}e^{V^\chi} \nn \\
e^{-iV_L}\hat{\bar{\nabla}}_+ e^{iV_L}\aleq \bbDB{+} \nn \\
e^{-iV_L}\hat{\nabla}_-e^{iV_L} \aleq \bbD{-}~.
\eer{lessnabla}
In this representation, the derivatives $\nabla$ are complicated.

The difference of two covariant derivatives is a
covariant tensor, and thus is a field-strength. We write four
spinor field-strengths that are the nonabelian generalizations of (\ref{lvmfield})\footnote{The signs in (\ref{gdef}) may seem inconsistent with hermitian conjugation; however, hermitian conjugation also changes the representation these Lie algebra-valued quantities act on, whereas in (\ref{gdef}) both $\bbG{}$ and $\bbGB{}$ act on the \textit{same} representation. This results in an extra $(-)$ sign.}:
\begin{eqnarray}\label{gdef}
&& \bbG{+}=i(\hat{\bar{\nabla}}_+ - \bar{\nabla}_+) ~~,~~~ \bbG{-}=i(\hat{\bar{\nabla}}_- - \bar{\nabla}_-) ~,\nn \\
&& \bbGB{+}=i(\hat{\nabla}_+ - \nabla_+) ~~,~~~ \bbGB{-}=i(\hat{\nabla}_- - \nabla_-) ~. \label{nonalvmfield}
\end{eqnarray}

We may shift the spinor covariant derivatives by these spinor field-strengths
as we wish; indeed, such shifts play a crucial role in understanding the kinetic terms of the large vector multiplet.

These spinor field-strengths obey (nonlinear) semichiral constraints. Using the identity
\beq
\{\hat{\bar\nabla}_\pm+\bar\nabla_\pm\, , \hat{\bar\nabla}_\pm-\bar\nabla_\pm\}=
\{\hat\nabla_\pm+\nabla_\pm\, , \hat\nabla_\pm-\nabla_\pm\}=0~,
\eeq{gid}
we find
\ber
(\hat{\bar\nabla}_\pm+\bar\nabla_\pm)\bbG{\pm}=0~&\Leftrightarrow&~
 \bar\nabla_\pm\bbG{\pm}-\ft{i}2\{\bbG{\pm},\bbG{\pm}\}=0~,~~\nn \\
(\hat\nabla_\pm+\nabla_\pm)\bbGB{\pm}=0
~&\Leftrightarrow&~\nabla_\pm\bbGB{\pm}-\ft{i}2\{\bbGB{\pm},\bbGB{\pm}\}=0~.
\eer{nonalvmsemi}

Higher dimension field-strengths may be found by taking anticommutators of covariant derivatives 
(\ref{chiralnabla}). In general, each chirality choice ((twisted)(anti)chiral) has three possible field-strengths and one trivial anticommutator. For example
\begin{equation}\label{dtriv}
 \{\nabla_+ , \nabla_- \} = 0~~,
 \end{equation}
 whereas
 \begin{equation}
\{\hat{\nabla}_+ , \nabla_- \}~~,~~~ \{\nabla_+ , \hat{\nabla}_- \}
  ~~,~~ \{\hat{\nabla}_+ , \hat{\nabla}_- \}
\end{equation}
are nonvanishing. Using (\ref{gdef}) and (\ref{dtriv}), we have:
\begin{equation}
\label{eq::FIELDSTRENGTH_RELATION}
 \{ \bbGB{+},\bbGB{-}\} =   \{\hat{\nabla}_+ , \nabla_- \} +  \{\nabla_+ , \hat{\nabla}_- \} - \{\hat{\nabla}_+ , \hat{\nabla}_- \}~.
\end{equation}

We thus find the independent field-strengths:
\begin{equation}
\label{naturals}
 \begin{array}{lll}
 \mbox{chiral:}
	&\myF=\{\hat{\bar{\nabla}}_+, \bar{\nabla}_- \} = 
	-i \bar{\nabla}_- \bbG{+}~,
	&\hat{\myF}= \{\bar{\nabla}_+, \hat{\bar{\nabla}}_- \} = 
	-i \bar{\nabla}_+ \bbG{-} \\
\mbox{antichiral:}
	&\bar{\myF}=\{\hat{\nabla}_+, \nabla_- \} = 
	-i \nabla_- \bbGB{+}~,
	&\hat{\bar{\myF}}=\{\nabla_+, \hat{\nabla}_- \} = 
	-i \nabla_+ \bbGB{-} \\
\mbox{twisted chiral:}
	&\tilde{\myF}=\{\hat{\bar{\nabla}}_+, \nabla_- \} = 
	i \hat{\bar{\nabla}}_+ \bbGB{-}~,
	&\hat{\tilde{\myF}} = \{\bar{\nabla}_+, \hat{\nabla}_- \} = 
	i \hat{\nabla}_- \bbG{+} \\
\mbox{twisted antichiral:}~
	&\bar{\tilde{\myF}} =  \{\hat{\nabla}_+, \bar{\nabla}_- \} = 
	i \hat{\nabla}_+ \bbG{-} ~,
	&\hat{\bar{\tilde{\myF}}}= \{\nabla_+, \hat{\bar{\nabla}}_- 
	\} = i \hat{\bar{\nabla}}_- \bbGB{+}
\end{array}
\end{equation}
The nonabelian field-strengths (\ref{naturals}) match, in the abelian case \cite{Lindstrom:2007sq}, with combinations of the form
\begin{equation}
 2 \ff = W+iB ~~,~~~ 2\ffh = W-iB~,  ~etc.
\end{equation}
Each field-strength has specific chirality properties that follow from its definition, \eg 
\begin{equation}\label{simplechiral}
\hat{\bar{\nabla}}_+ \ff = \bar{\nabla}_- \ff = 0~~,~~~
\bar{\nabla}_+ \ffh = \hat{\bar{\nabla}}_- \ffh = 0~,~etc.
\end{equation}

\subsection{Reduction to $\nll$ superspace}
As in the abelian case, we decompose 
\begin{equation}
 \bbD{\pm} = \ft{1}{2} (D_\pm -i Q_\pm)~,
\end{equation}
as well as
\begin{equation}
\left.\rm{Re}~ \bbG{\pm} \right|_{\nll} = \hee{\pm} ~~,~~~ 
\left.\rm{Im}~ \bbG{\pm} \right|_{\nll} = \hcc{\pm}~.
\end{equation}
The two sets of $\nll$ supercovariant derivatives decompose 
as\footnote{As is standard, in descending to
$\nll$ superspace, we reduce the gauge parameter 
to a single $\nll$ superfield; this means we perform
a partial (Wess-Zumino) gauge-fixing. As a result, 
$Q_\pm$ has no $\nll$ connection.}:
\begin{eqnarray}
\label{N11_DERIV}
&& \nabla_\pm = \ft{1}{2} (\mathcal{D}_\pm - i Q_\pm) ~~,~~~ 
\hat{\nabla}_\pm = \ft{1}{2}(\hat{\mathcal{D}}_\pm - i\hat{Q}_\pm) ~, \nn \\[3mm]
&&\hat{\mathcal{D}}_\pm = \mathcal{D}_\pm - 2i \hee{\pm} ~~,~~~ 
\hat{Q}_\pm = Q_\pm - 2i \hcc{\pm}~. 
\end{eqnarray}
The hatted set differs from the unhatted set by covariant field redefinitions; note that the redefinition exchanging $\mathcal{D}$ and $\hat{\mathcal{D}}$ is a  shift of the $\nll$ connections for $ \mathcal{D}_\pm = D_\pm + i A_\pm$ by $2i\hee{\pm}$. 

The field-strengths $F,\tilde F$ can be expressed in $\nll$ superspace by acting with $\hat{\mathcal{D}}_+,$ $\hat{Q}_+ ,$ $\mathcal{D}_-$ and $ Q_-$ on $\Xi^{1,2}_\pm$ 
({\it cf.} eq. (\ref{naturals})). 
We therefore define: 
\begin{equation}
\label{eq::check_derivatives}
 \check{\mathcal{D}}_+ = \hat{\mathcal{D}}_+ ~~,~~~  
 \check{Q}_+ = \hat{Q}_+ ~~,~~~  
 \check{\mathcal{D}}_- = \mathcal{D}_- ~~,~~~  
 \check{Q}_- = Q_-~.
\end{equation}
From the spinor derivatives (\ref{eq::check_derivatives}), we construct real $\nll$ scalars 
\begin{eqnarray}
\dq{\chi}=\{ \check{\mathcal{D}}_+ , \check{Q}_-\} \aleq -i \check{Q}_{(+}\hee{-)} + i \check{\mathcal{D}}_{[+} \hcc{-]}\nn \\
\dq{\phi} = \{ \check{\mathcal{D}}_- , \check{Q}_+\} \aleq i \check{Q}_{[+}\hee{-]} - i \check{\mathcal{D}}_{(+} \hcc{-)} \nn \\
\dq{\prime} = \{ \check{Q}_+ , \check{Q}_-\} \aleq i \check{Q}_{[+}\hcc{-]} + i \check{\mathcal{D}}_{(+} \hee{-)}\nn 
 \label{eq::unhatted_conventions}
\end{eqnarray}
as well as the field-strength
\begin{equation}
\label{eq::check_f}
 f\equiv \{ \check{\mathcal{D}}_+ , \check{\mathcal{D}}_-\} = i \check{Q}_{(+}\hcc{-)} + i \check{\mathcal{D}}_{[+}\hee{-]}
\end{equation}

These conventions simplify the $\nll$ reduction of all unhatted field-strengths 
by eliminating both $i\mathcal{D}\Xi$ and $\{ \Xi , \Xi\}$ terms:
\begin{eqnarray}
 \label{eq::CHIRAL_FIELD_STRENGTHS_REDUCTION}
4\ff| \aleq
		f - \dq{\prime} + 
		i (\dq{\chi}+\dq{\phi}) 
		\\
4\ffb| \aleq
		f - \dq{\prime} -
		i (\dq{\chi}+\dq{\phi}) \nn \\
4\fft| \aleq
		f + \dq{\prime} -
		i (\dq{\chi}-\dq{\phi}) \nn \\ 
4\fftb| \aleq
		f + \dq{\prime} +
		i (\dq{\chi}-\dq{\phi}) \nn \\
\ffh| \aleq
		\fftb - i \check{\mathcal{D}}_+ \hee{-} +
		\{ \hee{+},\hee{-}\} - \{ \hcc{+},\hcc{-}\} -
		i(i\check{\mathcal{D}}_+ \hcc{-} - 
		\{ \Xi^{(1}_+ , \Xi^{2)}_- \}) \nn \\
\ffhb| \aleq
		\fft - i \check{\mathcal{D}}_+ \hee{-} +
		\{ \hee{+},\hee{-}\} - \{ \hcc{+},\hcc{-}\} +
		i(i\check{\mathcal{D}}_+ \hcc{-} - 
		\{ \Xi^{(1}_+ , \Xi^{2)}_- \}) \nn \\
\ffth| \aleq
		\ff + i \check{\mathcal{D}}_- \hee{+} +
		\{ \hee{+},\hee{-}\} + \{ \hcc{+},\hcc{-}\} +
		i(i\check{\mathcal{D}}_- \hcc{+} - 
		\{ \Xi^{[1}_+ , \Xi^{2]}_- \}) \nn \\
\fftbh| \aleq
		\ffb + i \check{\mathcal{D}}_- \hee{+} +
		\{ \hee{+},\hee{-}\} + \{ \hcc{+},\hcc{-}\} -
		i(i\check{\mathcal{D}}_- \hcc{+} - 
		\{ \Xi^{[1}_+ , \Xi^{2]}_- \}) \nn \\
\end{eqnarray}

The $\nll$ fields (\ref{eq::check_derivatives}) and (\ref{eq::check_f}) could be redefined \cite{ryb} by real shifts of 
$\dq{i}$ and the connections $A_\pm$. The redefinitions
\begin{equation}
 \label{eq::hatted_conventions}
\dqh{\chi}=\{ \mathcal{D}_+ , \hat{Q}_-\} ~~,~~~
\dqh{\phi} = \{ \hat{\mathcal{D}}_- , Q_+\} ~~,~~~
\dqh{\prime} = \{ Q_+ , \hat{Q}_-\} ~~\mbox{and} ~~ 
\hat{f} = \{ \mathcal{D}_+ , \hat{\mathcal{D}}_-\} 
\end{equation}
simplify the reduction to $\nll$ for the 
field-strengths $\ffh,\ffth$ and introduces extra 
$i\mathcal{D}\Xi$ and $\{\Xi,\Xi\}$ terms into 
the reduction for the field-strengths $\ff,\fft$.

An immediate consequence of the structure of these definitions is that we are able to 
remove both $i\mathcal{D}\Xi$ and $\{\Xi,\Xi\}$ terms from either the hatted 
set or the unhatted set of $\nZZ$ field-strengths, but not both simultaneously. 
This result greatly simplifies our discussion of actions for the large vector multiplet.

\subsection{The action in $\nll$ superspace}
\subsubsection{Generalities}
We descend to $\nll$ superspace by rewriting the measure in terms of  $D_\pm,Q_\pm$ and
explicitly evaluating the $Q$ derivatives (see, \eg \cite{Lindstrom:2007sq}). 
Starting with the $\nZZ$ superspace measure and an $\nZZ$ Lagrange density $K$, we write an action
\begin{equation}
 S = \int d^2 \xi D_+ D_- Q_+ Q_- ~ K~.
\end{equation}
Since $K$ is a gauge scalar, we are free to choose whether
$Q(D)$-derivatives acts on $K$ as $Q(\mathcal{D})$ or $\hat{Q}(\hat{\mathcal{D}})$.

This leads to a subtlety in descending to $\nll$ superspace.  Unlike the abelian case, where one can simply exchange $Q$ and $D$-derivatives using the complex structure
\begin{equation}
\label{eq::simple_reduction}
 Q_\pm \varphi^i =   \jj{\pm}{i}{j}D_\pm \varphi^j ~,
\end{equation}
the natural nonabelian field-strengths (\ref{naturals}) have chirality properties with respect to different $\nZZ$ supercovariant derivatives. This results in a possible shift of the relation (\ref{eq::simple_reduction}) with a spinor multiplet $\bbG{}$. For example, the action of $\check{Q}_\pm$ on the (anti)chiral field-strengths reads
\begin{equation}
\label{eq::Q_ACTION}
 \begin{array}{c||c|c}
~	& \check{Q}_+			& \check{Q}_- \\ \hline \hline
\ff	&  i \check{\mathcal{D}}_+ \ff| 
	&  i \check{\mathcal{D}}_- \ff|\\
\ffb	& -i \check{\mathcal{D}}_+ \ffb| 
	& -i \check{\mathcal{D}}_- \ffb|\\
\ffh	& ~~~i \check{\mathcal{D}}_+ \ffh| - 2 [\bbG{+} ,\ffh]|
	& ~~~i \check{\mathcal{D}}_- \ffh| + 2 [\bbG{-} ,\ffh]|\\
\ffhb	& -i \check{\mathcal{D}}_+ \ffhb|  + 2 [\bbGB{+}, \ffhb]|
	& -i \check{\mathcal{D}}_- \ffhb|  - 2 [\bbGB{-} ,\ffhb]| 
\end{array}
\end{equation}
For the action of $\check{Q}_+ \check{Q}_-$ we find:
\begin{equation}
\label{eq::QQ_ACTION}
 \begin{array}{c||c}
~	& \check{Q}_+ \check{Q}_-\\ \hline \hline
\ff	& \check{\mathcal{D}}_- \check{\mathcal{D}}_+ \ff| 
+i [\dq{\phi} , \ff]|\\
\ffb	& \check{\mathcal{D}}_- \check{\mathcal{D}}_+ \ffb| 
-i [\dq{\phi},\ffb]|\\
\ffh	& - \{ i \check{\mathcal{D}}_- + 2 \bbG{-}, [i \check{\mathcal{D}}_+ - 2\bbG{+},\ffh]\} + [f+\dq{\prime}+i\dq{\chi}-2i\check{\mathcal{D}}_+ \bbG{-},\ffh]|\\
\ffhb	& - \{ i \check{\mathcal{D}}_- + 2 \bbGB{-}, [i \check{\mathcal{D}}_+ - 2\bbGB{+},\ffhb]\} - [f+\dq{\prime}-i\dq{\chi}-2i\check{\mathcal{D}}_+ \bbGB{-},\ffhb]|
\end{array}
\end{equation}
where we used the anticommutators
\begin{eqnarray}
 \{ \check{Q}_+ , i\check{\mathcal{D}}_- + 2 \bbG{-} \} \aleq f + \dq{\prime} + i \dq{\chi} - 2i \check{\mathcal{D}}_+ \bbG{-} \nn \\
 -\{ \check{Q}_+ , i\check{\mathcal{D}}_- + 2 \bbGB{-} \} \aleq f + \dq{\prime} - i \dq{\chi} - 2i \check{\mathcal{D}}_+ \bbGB{-} 
\end{eqnarray}

\subsubsection{Evaluating the actions}

Physically sensible actions cannot have terms with higher derivatives on fermions. We now generalize the results presented in \cite{ryb}, where field redefinitions were found that allowed us to write down actions for (anti)chiral field-strengths and twisted (anti)chiral field-strengths.

Na\"ively, quadratic terms in (anti)chiral field-strengths may appear in three flavors:  $\ff \ffb$, $\ffh\ffhb$, 
and $\ff \ffhb + c.c$. Using the results of the previous section, one can show that there exist field redefinitions
that eliminate higher derivative terms for \textit{either} of the first two but not the last $\nZZ$ action. Extending these results to the twisted (anti)chirals we find that sensible gauge actions in $\nZZ$ are either combinations of only hatted field-strengths or only unhatted ones.

From (\ref{naturals}) and (\ref{simplechiral}), we see that \eg the 
action\footnote{It is essential to have both $\ff \ffb$ and $\fft \fftb$ 
terms in the $\nZZ$ Lagrange-density to give dynamics all $\nll$ 
multiplets. In the abelian case, where terms of the form $\{ \Xi ,\Xi \}$ 
vanish, actions that contain only chiral or only twisted chiral multiplets are also possible. 
These actions are discussed in \cite{ryb}.} $K=\mbox{Tr}(\ff \ffb - \fft \fftb)$ 
is conveniently reduced to $\nll$ superspace by acting with $Q$-derivatives 
as $Q_-$ and $\hat{Q}_+$:
\begin{eqnarray}
\label{eq::n11_action}
 \int D_+ D_- Q_+ Q_- ~ \mbox{Tr} \big(\!\!\!\!\!\!&&\!\!\!\!\ff \ffb-\fft \fftb)  \\
=\frac14\int D_+ D_- ~ \mbox{Tr} \big(\!\!\!\!\!\!&&\!\!\!\! \check{\mathcal{D}}_+(f-\dq{\prime})\check{\mathcal{D}}_- (f-\dq{\prime})+\check{\mathcal{D}}_+ (\dq{\chi}+\dq{\phi}) \check{\mathcal{D}}_- (\dq{\chi}+\dq{\phi})+ \nn \\
\!\!\!\!\!\!&&\!\!\!\!\check{\mathcal{D}}_+ (f+\dq{\prime})\check{\mathcal{D}}_- (f+\dq{\prime})+\check{\mathcal{D}}_+(\dq{\chi}-\dq{\phi}) \check{\mathcal{D}}_-(\dq{\chi}-\dq{\phi})-
2[\dq{\prime},\dq{\chi}]\dq{\phi} \big)\nn \\
=\frac12\int D_+ D_- ~ \mbox{Tr} \big(\!\!\!\!\!\!&&\!\!\!\! \check{\mathcal{D}}_+f\check{\mathcal{D}}_- f+\check{\mathcal{D}}_+ \dq{\chi} \check{\mathcal{D}}_- 
\dq{\chi}+ \check{\mathcal{D}}_+ \dq{\prime}\check{\mathcal{D}}_-\dq{\prime}+\check{\mathcal{D}}_+\dq{\phi} \check{\mathcal{D}}_-\dq{\phi}-
[\dq{\prime},\dq{\chi}]\dq{\phi} \big)\nn~.
\end{eqnarray}
It is interesting to notice that the action (\ref{eq::n11_action}), and  in particular the scalar commutator term, is reminiscent of the action for $N=2~d=4$ super Yang-Mills theory. This suggests that the large vector multiplet action has $N=(4,4)$ supersymmetry. We leave this for future work.

Other possible contributions to the nonabelian large vector multiplet action originate from superpotentials which are encoded in four complex functions. Their reduction to $\nll$ superspace reads:
\begin{equation}
 \label{eq::superpotential}
 S_{sp} = 2 \left.\mbox{Re}  \int D_+D_-  \left( P(\ff) + \tilde{P}(\fft) + \hat{P}(\ffh)+\hat{\tilde{P}}(\ffth) \right)\right|~.
\end{equation}
The criterion for consistent superpotential terms, which is the absence of terms of the form $\mathcal{D}_+ \Xi_- \mathcal{D}_- \Xi_+ $, is automatically met for {\em any} of the field redefinition required to make the 
kinetic terms consistent. In the abelian limit, terms involving $\ff\ffh$ are also chiral, but were excluded by a consistency condition found in \cite{ryb}; here, these terms are not chiral and hence are automatically excluded.

The superpotential terms (\ref{eq::superpotential}) give mass terms for the scalar multiplets as well actions for the spinor multiplets $\Xi_\pm^{1,2}$ (as in the abelian case \cite{ryb}).

\Section{Nonabelian Semichiral Vector Multiplet}
The strategy we follow for the semichiral vector multiplet is very similar to the one we use for the large vector multiplet. However,
since the gauge parameters are semichiral, we find unique gauge covariant spinor
derivatives and all the field-strengths arise in the usual way as (anti)commutators.

We take the nonabelian generalization of (\ref{svm} , \ref{semiv} ,  \ref{semivt}) to be:
\ber
g(\lam_L,\lam_R) {e^{i\vv{} }} &=& e^{i\Lambda_L} e^{i\vv{} } e^{-i\Lambda_R}\nn\\
g(\lam_L,\lamb_R){e^{i\vvt{}}} &=& e^{i\Lambda_L} e^{i\vvt{} } e^{-i\bar\Lambda_R}\nn\\
g(\lam_L){e^{\vv{L}}} &=& e^{i\bar\Lambda_L}e^{\vv{L}} e^{-i\Lambda_L}\nn\\
g(\lam_R){e^{\vv{R}}} &=& e^{i\bar\Lambda_R} e^{\vv{R}} e^{-i\Lambda_R}~.
\eer{nonasemi}
As for the large vector multiplet, not all of these potentials are independent. We impose the
gauge covariant constraint 
\ber
 e^{i\vv{}} = e^{-\vv{L}} e^{i\bar{\vvt{}}}
\eer{vvtsemicon}
as well as the  gauge covariant reality constraint
\ber
e^{i\vv{}} = e^{-\vv{L}} e^{i\bar\vv{}} e^{\vv{R}}~.
\eer{gcovsemi}
Again, the coupling to (semichiral) matter is straightforward:
\ber	
&&K(\bbXB{L}\cdt \bbX{L} , \bbXB{R}\cdt \bbX{R} , 
\bbXB{L}\cdt \bbX{R} , \bbXB{R}\cdt \bbX{L} ) \to 
\nonumber \\
&&\qquad\qquad K(\bbXB{L}{e^{\vv{L}}\!}\cdt \bbX{L} , 
\bbXB{R}{e^{\vv{R}}\!}\cdt \bbX{R} ,  
\bbXB{L} e^{i\bar{\vvt{}}\!}\cdt \bbX{R}, \bbXB{R} e^{-i\vvt{}\!}\cdt \bbX{L} )~.  
\eer{GaugeSemi}

The covariant derivatives read (in the left semichiral representation):
\ber
\bar{\nabla}_+ \aleq \bbDB{+} \nn \\
\bar{\nabla}_- \aleq e^{i\vv{}} \bbDB{-} e^{-i\vv{}} \nn \\
\nabla_+ \aleq e^{i\vv{}}e^{i\vvtb{}} \bbD{+} e^{-i\vvtb{}} e^{-i\vv{}}  = e^{i\vvt{}}e^{-i\vvb{}} D_+ e^{i\vvb{}} e^{-i\vvt{}}  \nn = e^{-\vv L}D_+ e^{\vv L}
\\
\nabla_- \aleq e^{i\vvt{}} \bbD{-} e^{-i\vvt{}}
\eer{semichiral_der1}

The nonabelian generalization of the field-strengths (\ref{semifield}) is
\begin{eqnarray}
\ffs = i \{ \bar{\nabla}_+ , \bar{\nabla}_- \} &~& \ffbs = -i \{ \nabla_+ , \nabla_-\} \nn\\
\ffts = i \{ \bar{\nabla}_+ , \nabla_- \} &~& \fftbs = -i \{ \nabla_+ , \bar{\nabla}_-\}~,
\end{eqnarray}
which are covariantly chiral and twisted chiral scalars.

\subsection{Reduction to $\nll$ superspace}
Having a single set of $\nZZ$ supercovariant derivatives, we introduce the $\nll$ derivatives
\begin{equation}
 \nabla_{\pm} = \ft{1}{2}( \mathcal{D}_\pm - iQ_\pm )~,
\end{equation}
that give the $\nll$ field-strength for the connections
\begin{equation}
f= \{ \mathcal{D}_+ , \mathcal{D}_- \} = \{\nabla_+ + \bar{\nabla}_+ , \nabla_- + \bar{\nabla}_- \} = -i \left. \left( \ffs+ \ffts - \ffbs - \fftbs \right) \right|~,
\end{equation}
and three scalars that follow the notation of \cite{Lindstrom:2007vc}
\beq
\dt1=\!\left.\left(\ffs+\ffbs\right)\right|~~,~~~\dt2=\!\left.\left(\ffts+\fftbs\right)\right|
~~,~~~\dt3=\!\left.i\!\left(\ffs-\ffbs-\ffts+\fftbs\right)\right|~.
\eeq{ffN11}

In \cite{Lindstrom:2007vc} we have obtained the $D$-term action, which is a simple sum of kinetic terms for the (twisted) chiral field-strengths. 
Using the chirality properties of the field-strengths
\begin{equation}
 Q_\pm (\ffs,\ffbs,\ffts,\fftbs) = \mathcal{D}_\pm (i\ffs,-i\ffbs,\pm i \ffts, \mp i \fftbs) ~,
\end{equation} 
the nonabelian action
\begin{equation}
S_\mathbb{X} =  \int d^2 \xi\, D_+ D_- Q_+ Q_- 
\mbox{Tr} \left( \ffs\ffbs-\ffts\fftbs \right)
\end{equation}
reduces to the $\nll$ action
\begin{eqnarray}
S_{\mathbb{X}} = \int d^2 \xi  D_+ D_- ~ \mbox{Tr} \!\!\! &&\!\!\! \left( \ft{1}{2} \mathcal{D}_+ f\mathcal{D}_- f + \ft{1}{2} \mathcal{D}_+ \dt{3} \mathcal{D}_- \dt{3} \right. \nn \\
\!\!\! && \left.+  \mathcal{D}_+ \dt{1} \mathcal{D}_- \dt{1} + \mathcal{D}_+ \dt{2} \mathcal{D}_- \dt{2} - [\dt{1},\dt{2}]\dt{3} \right) ~.
\label{Ssemivec11}
\end{eqnarray}

In a similar fashion, we write the superpotential terms in $\nll$ superspace:
\begin{equation}
S_{P} = 2\left. \mbox{Re} \int  D_+ D_- \left( P_1 (\ffs) +   P_2 (\ffts) \right)\right|~.
\end{equation}

These terms include mass and Fayet-Illiopoulos terms.

\section{Conclusion}
In this paper we have extended the results of \cite{Lindstrom:2007vc,Gates:2007ve,ryb} 
to the nonabelian case. While the semichiral vector multiplet generalizes straightforwardly,
the extension for the large vector multiplet gave rise to subtleties and ambiguities 
that were not present in the abelian case, namely, the different chirality conditions  (\ref{naturals}) 
that follow from the doubled set of supercovariant derivatives $\nabla , \hat{\nabla}$.

The nonabelian extension sheds light on the origins of some of the constraints on actions for the large vector multiplet \cite{ryb}. In particular, in the nonabelian case, $D$-terms are further restricted to four possible kinetic terms and the restrictions on superpotential terms found in \cite{ryb} are an immediate consequence of the incompatibility in chirality properties for $\nZZ$ field-strengths. 

Applications of our work are both geometric and physical: They should give us insight into nonabelian
generalized K\"ahler quotients, and allow the construction of new generalized gauged linear sigma 
models with NS-NS two-form flux.

\bigskip\bigskip
\noindent{\bf\Large Acknowledgement}:
\bigskip\bigskip

\noindent
We are grateful to the 2007 and 2008 Simons Workshops for providing
the stimulating atmosphere where part of this work was carried out. 
UL supported by EU grant (Superstring theory)
MRTN-2004-512194 and VR grant 621-2006-3365.
The work of MR and IR was supported in part by NSF grant no.~PHY-0653342.
The research of R.v.U. was supported by
Czech ministry of education contract No.~MSM0021622409.
The research of M.Z. was
supported by VR-grant 621-2004-3177.

\end{document}